**Title**

Non-invasive super-resolution imaging through scattering media using fluctuating speckles

**Authors**

Xiangwen Zhu[1], Sujit Kumar Sahoo[2†], Giorgio Adamo[3], Landobasa Y.M. Tobing[1], Dao Hua Zhang[1], Cuong Dang[1†]

**Affiliations**

[1]Centre for Optoelectronics and Biophotonics (COEB), School of Electrical and Electronic Engineering, The Photonics Institute (TPI), Nanyang Technological University, 50 Nanyang Avenue, Singapore 639798, Singapore.

[2]School of Electrical Sciences, Indian Institute of Technology Goa, Goa 403401, India.

[3]Centre for Disruptive Photonic Technologies, SPMS, TPI, Nanyang Technological University, Singapore 637371, Singapore.

[†]Corresponding author. Email: sujit@iitgoa.ac.in (S.K.S); HCDang@ntu.edu.sg (C.D.)

**Abstract**

Extending super-resolution imaging techniques to objects hidden in strongly scattering media potentially revolutionize the technical analysis for much broader categories of samples, such as biological tissues. The main challenge is the media's inhomogeneous structures which scramble the light path and create noise-like speckle patterns, hindering the object's visualization even at a low-resolution level. Here, we propose a computational method relying on the object's spatial and temporal fluctuation to visualize nanoscale objects through scattering media non-invasively. The fluctuating object can be achieved by random speckle illumination, illuminating through dynamic scattering media, or flickering emitters. The optical memory effect allows us to derive the object at diffraction limit resolution and estimate the point spreading function (PSF). Multiple images of the fluctuating object are obtained by deconvolution, then super-resolution images are achieved by computing the high order cumulants. Non-linearity of high order cumulant significantly suppresses the noise and artifacts in the resulting images and enhances the resolution by a factor of $\sqrt{N}$, where N is the cumulant order. Our non-invasive super-resolution speckle fluctuation imaging (NISFFI) presents a nanoscopy technique with very simple hardware to visualize samples behind scattering media.

**Teaser (125 characters)**

A computational approach relied on fluctuating speckles to non-invasively visualize objects behind strong scattering media at sub-wavelength scale.

**Introduction**

Most biological tissues are strongly scattering media, which scramble light propagation paths, posing a considerable challenge for imaging (Fig. 1A). A camera typically captures just speckle images, which are seemingly random noise. Various techniques based on the optical memory effect (*1, 2*) in thin scattering media have recently been introduced to overcome this challenge. The optical memory effect implies that the speckles' autocorrelation is similar to the object's autocorrelation, allowing to non-invasively derive the object by a phase retrieval algorithm – an iterative optimization approach (*3-5*). The object can also be more deterministically computed using a deconvolution algorithm (*6-8*) if the media's point spreading function (PSF), which is typically measured by an invasive approach, is known. By multiplexing multiple uncorrelated PSFs in a single speckle image, it is possible to achieve single-shot multispectral (*9*) or multi-view imaging (*10*) with the help of scattering media. These snapshot approaches have successfully recovered the object, but the best achievable resolution is still the diffraction limit (Fig. 1B), $0.61\lambda/NA$, where $\lambda$ is wavelength and NA is the numerical aperture of imaging optics (*11*).

Imaging beyond the optical diffraction limit is mature, and has been revolutionizing technical analysis for transparent samples in various fields from biological to physical sciences. Several examples are either using single fluorescence emitters like stochastic optical reconstruction microscopy, STORM(*12, 13*), and photo-activated localization microscopy, PALM(*14*), or sharpening a known point spreading function (PSF) such as stimulated emission depletion, STED(*15*), non-linear structure illumination microscopy, SSIM, (*16*) and super-resolution optical fluctuation imaging, SOFI (*17*). The significance of these optical nanoscopy techniques is to provide detailed information at electron microscopes' resolution in alive samples such as cells or transparent biological tissues(*18, 19*). The transparency requirement for optical imaging often needs to engage advanced techniques to engineer the biological tissues, such as primate-optimized uniform clearing method (PuClear) for only in-vitro samples, unfortunately (19). Expanding the capability of super-resolution imaging to translucent samples or scattering media will open many opportunities for not only biological applications but also various applications in science and technology. Recently, Wang et al. introduced a stochastic optical scattering localization imaging (SOSLI) technique (*11*) that utilizes speckle correlation to image non-invasively through strongly scattering media at 100-nm resolution. Although the SOSLI's resolution limit is merely governed by signal to noise ratio (SNR), the requirement on the sparse individual blinking emitters - only one emitter per diffraction-limited area is on at a time - poses considerable challenges for current quantum or molecular emitters to achieve a practical SNR.

In this work, we present a speckle fluctuation imaging method, which allows non-invasive visualization of objects behind scattering media with super-resolution. The technique relies on intensity fluctuation from an object and its speckle correlation, thanks to the optical memory effect of scattering media. Unlike SOSLI, our hidden object can have denser emitters per diffraction-limited area to boost the SNR. We also can illuminate a transmission sample (no emitters) with an externally moving point source (Fig. 1E), where high SNR can be achieved merely by utilizing a stronger illumination source. The object's intensity fluctuates spatially and temporally (Fig. 1C), i.e. uncorrelated in time and space. A camera captures multiple fluctuating speckle frames, which allow us to recover multiple images of the uncorrelated fluctuating object. A super-resolution image is achieved by calculating the high-order cumulants (Fig. 1D). The simulation and experimental results show that our non-invasive super-resolution speckle fluctuation imaging (NISSFI) can break diffraction-limit and achieve the resolution improvement by a factor of $\sqrt{N}$, where N is the cumulant order. We experimentally demonstrate our NISSFI's resolution at 300 nm and FWHM of 257 nm with NA = 0.37 (the theoretical diffraction limit of 875 nm). NISSFI can also be used in an adaptive mode to image through dynamic scattering media by utilizing a small correlation (0.16) of the media between two consecutive shots.

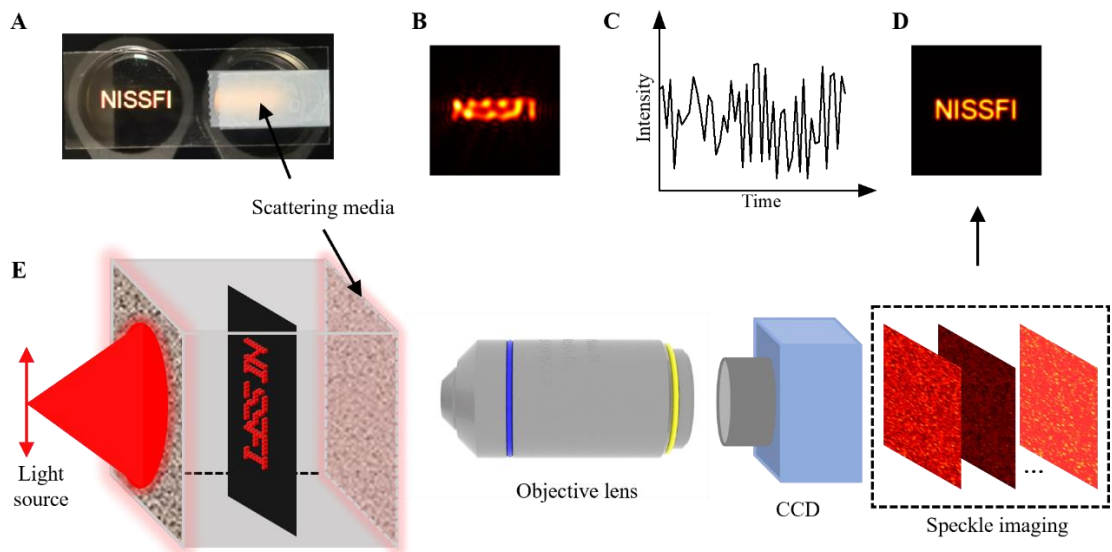

**Fig. 1. Schematic of NISSFI.** (**A**) The scattering media scramble light path causing object information loss in comparison to transparent media. (**B**) The recovered image with diffraction-limit. (**C**) The speckle illumination intensity is fluctuating temporally due to the moving point source. (**D**) The recovered super-resolution image by NISSFI. (**E**) A pseudo-thermal point source is moving outside the front scattering media, forming time-varying speckle illumination in object plane, and light from object is scattered by back scattering media, resulting in a sequence of speckle fluctuation frames in the CCD.

## Results

### Principle and simulation of NISSFI

Our principle and simulation are shown in Fig. 2. A sample consists of several intensity-fluctuating dots with different inter-distances (300 nm, 250 nm, and 200 nm). The size of each dot is one pixel of 25 nm. When passing through scattering media, light from the sample forms speckle frames that are convolution of the samples with a PSF thanks to optical memory effect (*3*). A sequence of fluctuating speckle frames is captured as the dots are fluctuating (Fig. 2A). When the number of frames is large enough, the averaged frame is considered as the speckle frame from the fully bright object. A phase-retrieval (PR) algorithm, which is a mature technique in previous studies (*3, 4, 11*), is used to recover the fully bright object (Fig. 2B). Certainly, the reconstruction has artifacts and cannot break the diffraction-limit 342 nm, which is simulated for wavelength $\lambda = 532$ nm and NA = 0.95, defined by the aperture on scattering media and distance from object to scattering media. We then can estimate the PSF from deconvolution of the averaged speckle frame with the recovered object. In the simulation with SNR of 40 dB, our estimated PSF (Fig. 2B) can have a very high correlation (exceeding 0.90) with the actual PSF (Fig. 2A). With the estimated PSF, a series of fluctuating object's images (Fig. 2B) are derived by deconvolution from corresponding fluctuating speckle frames. It is worth noting that PR results are blind to absolute position and central flip. While the absolute position is not essential as it only affects the final image's position, the orientation information is crucial to estimate the PSF. Only one central flip orientation allows the sub-sequence deconvolution to successfully derive a series of fluctuating object frames (Fig. 2B). We have to try two possibilities and choose the correct one.

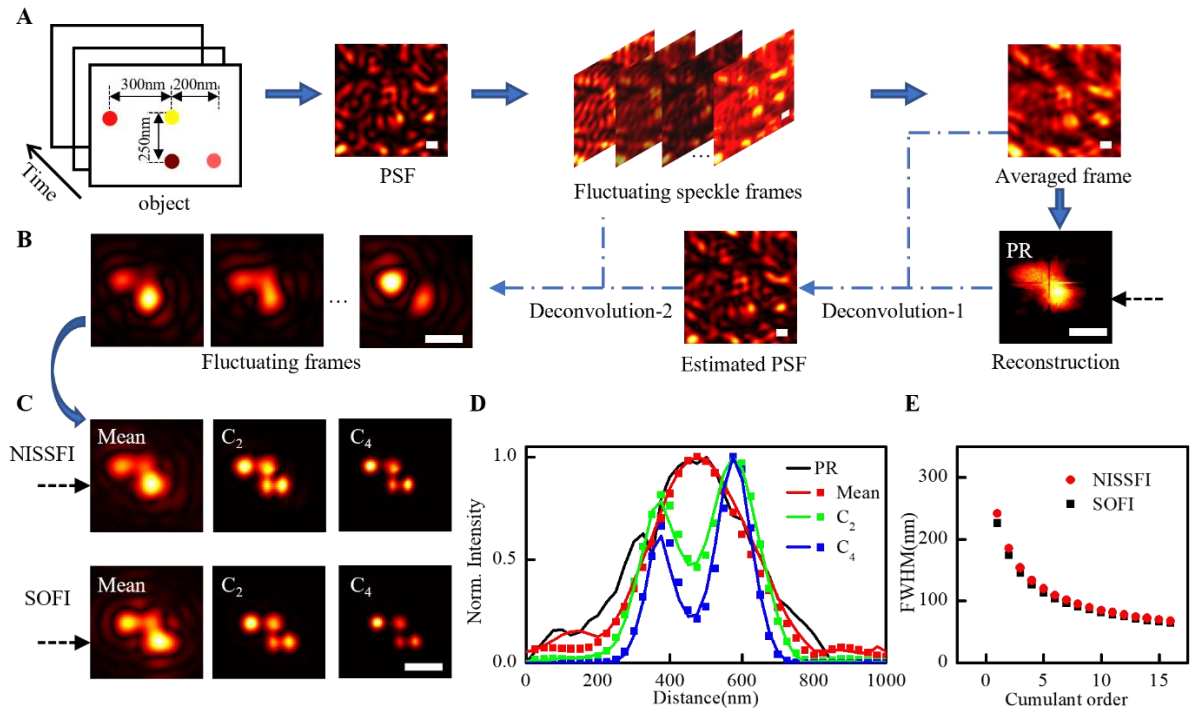

**Fig. 2. Principle and simulation results.** (**A**) A fluctuating object is convolved with a PSF, resulting a sequence of fluctuating speckle frames. Their averaged frame can be utilized to reconstruct the object by Phase-Retrieval algorithm (PR). (**B**) The estimated PSF by deconvolution of the averaged frame with the reconstructed image. A sequence of fluctuating frames can be derived from the estimated PSF and fluctuating speckle frames by deconvolution. (**C**) The first row is NISSFI results (Mean image, $C_2$: 2nd order cumulants, $C_4$: 4th order cumulants) from fluctuating frames in (**B**); The second row is classical SOFI technique where the object is imaged by lens without scattering media. (**D**) The intensity profile along the line indicated by the dashed arrows in (**C**) and (**B**). Line indicates NISSFI. Square indicates classical SOFI. Scale bar: 500 nm.

The series of fluctuating object's images recovered above are of low resolution (diffraction-limited) and exhibit some artifacts. However, the higher-order cumulants of these images can break diffraction-limit and remove artifacts as shown in Fig. 2C. The first order cumulant image (i.e., the mean image) is blurry. We can barely resolve 300 nm gap between two dots. However, after subtracting the mean, we calculate higher order cumulants and reveal object much clearer. The second order cumulant ($C_2$) and the 4th order cumulant ($C_4$) break the diffraction-limit of 342 nm, resolving the 250 nm and 200 nm gap very well. More interestingly, a lot of artifacts in deconvolution images and also in the PR image are significantly suppressed as the result of the non-linear effect in high order cumulants. We also simulate the conventional SOFI, where the fluctuating object is imaged by an objective of equivalent NA (0.95) in a transparent environment (i.e., no scattering media). The lower row in Fig. 2C shows comparable resolution of conventional SOFI with our NISFFI. The intensity along the horizontal line indicated by dashed arrows in Fig. 2B-C shows the significant resolution improvement for both techniques as expected (Fig. 2D). Fig. 2E presents the full width at half maximum (FWHM) of the point as a function of cumulant order. While the SOFI is marginally better, both techniques show the FWHM reduction by a factor of $\sqrt{N}$, where $N$ is cumulant order. The FWHM of our NISSFI can reach 242 nm and 186 nm for $C_2$ and $C_4$ image, respectively. A resolution of 100 nm or higher can be achieved with cumulants order greater than 8. By estimating the PSF non-invasively, the scattering media become scattering lens, allowing us to utilize conventional super-resolution techniques and achieve super-resolution imaging through scattering media. More

Page 4 of 17

interestingly, the non-linearity in NISSFI helps remove lot of artifact in PR and deconvolution imaging results.

**NISSFI with speckle illumination**

The key requirement for NISSFI is the intensity fluctuation of the points that comprise the object. We first achieve this with incoherent speckle illumination as shown in Fig. 1A, where a point source is moving parallelly to the scattering media. Light passing through the first scattering media creates speckle patterns on the sample, which is a thick metal film with a transmission pattern made by a focus ion beam (FIB) technique(*20*). The moving point source creates randomly changing speckle illumination, making a temporally and spatially fluctuating objects in the transmission mode. The speckle size of the illumination pattern needs to be as small as possible to maintain the independent fluctuation of the points on object. We focus a laser beam tightly into the surface of a rotating diffuser to create a pseudo-incoherent point source and utilize a thick scattering media (5mm thick polyurethane foam) to achieve our speckle size of about 700 nm. Thicker scattering media allow larger scattering angles, i.e., higher spatial frequency (smaller size) for illumination speckle pattern. However, thicker media reduce the fluctuation effect due to incoherent nature of the light source. Note that fluctuating illumination speckle patterns can be achieved by utilizing dynamic scattering media instead of moving the point source.

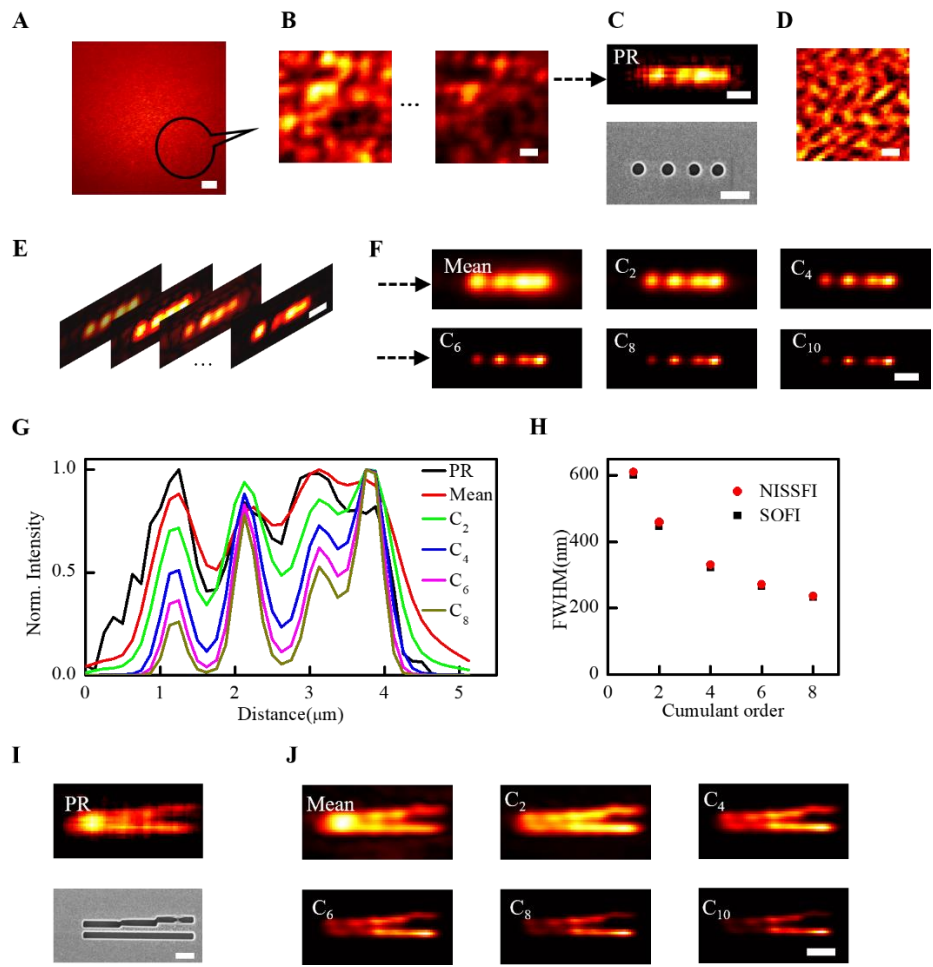

**Fig. 3. Experiment resolution improvement.** (**A**) A typical speckle frame. (**B**) Zoom in the same position of different fluctuating speckle frames, similar structure but different details. (**C**) PR reconstruction and SEM image. (**D**) Estimated PSF. (**E**) A sequence of deconvolution frames. (**F**) NISSFI results: Mean, $2^{nd}$, $4^{th}$, $6^{th}$, and $8^{th}$ order cumulants. (**G**) The intensity profile along the line indicated by dashed arrow in (**F**) and (**C**): black-PR; red-Mean; green-$C_2$; blue-$C_4$; pink-$C_6$; dim yellow-$C_8$. (**H**) The FWHM decrease with cumulant order increase (**I**) PR reconstruction and SEM image of continuous sample. (**J**) NISSFI results of continuous sample. Scale bar: 25 μm in (**A**) and 1.0 μm in others.

Light from the fluctuating object passing through the second scattering media and captured by an objective lens and a camera. A typical speckle frame is presented in Fig. 3A with high contrast and a large SNR. Fig. 3B shows examples of the same small area (dashed circle in Fig. 3A) in 2 fluctuating speckle frames, highlighting the generally similar structure with fluctuation speckle intensity. The averaging frame, which provides higher SNR signal, is then used for the PR reconstruction. The PR result and ground truth (SEM image) are shown in Fig. 3C. The former is poorly reconstructed with excessive noise and artifacts, but still allow us to estimate PSF (Fig. 3D). Then we derive a series of fluctuating object' images (Fig. 3E) by deconvolution from the fluctuation speckle frames. Certainly, these derived images are low resolution with some artifacts. For comparison, the NISSFI results in Fig. 3F show super-resolution especially at higher order cumulants. The mean image can resolve 3 dots as the PR reconstruction image, but it is a lot smoother and less artifact because of deconvolution process and averaging over a large number of images. The four dots are resolved gradually in higher order cumulants. The 400 nm diameter dots, with the narrowest gap of 400 nm are observed. In Fig. 3G, we plot the normalized intensity across the line indicated by dashed arrows in Fig. 3C and Fig. 3F. Obviously, the intensity profiles of PR results (black) and mean results (red) do not really show two dots on the right. However, these two peaks on the right gradually become distinguishable at higher-order cumulants. The FWHM of the dot's intensity at different cumulant orders is shown in Fig. 3H. We also remove the scattering media to do the conventional SOFI, where the objective lens captured diffraction limited images of fluctuating object directly (Supplementary Fig. S1). In fact, the NISSFI results are slightly better than direct imaging with microscope objective because the scattering media is closer to the sample than the objective therefore having higher NA for imaging(7). Both results are consistent with the $\sqrt{N}$ improvement as presented in simulation above. We can certainly calculate higher cumulant order and achieve the FWHM as a single pixel size (125 nm in our case). Figure 3I-J present NISSFI results for continuous samples (as opposed to discrete dots) where the line width is 300 nm and the vertical gap between lines are 300, 400, 600 nm. We also present the NISSFI results for another sample with the linewidth of 200 nm in Supplementary Fig. S2. Again, imaging results at higher-order cumulants are sharper and break the diffraction limit.

**NISSFI with speckle illumination through dynamic scattering media**

A fluctuation object can also be achieved by a static point source illuminating through dynamic scattering media, which are naturally happened in most of the practical applications. However, dynamic scattering media imply a dynamic PSF as well, and our static PSF approach in Fig. 3 will not work. We propose here an adaptive approach utilizing the robustness of deconvolution method and our ability to control the illumination source to do NISSFI through dynamic scattering media.

A proof-of-concept simulation for NISSFI through dynamic scattering media with 20dB SNR is shown in Fig. 4. First, we increase the size of illumination source so that the hidden object is illuminated fully by uniform light, i.e., no speckle illumination, no fluctuation. This can be done by simply defocusing the laser beam on the rotating diffuser in the previous experiment. Then we immediately illuminate the sample with the smallest point source (tightly focus laser beam on the rotating diffuser). The two speckle frames (a frame pair) captured at two deferent times present some decorrelation due to dynamic scattering media. The process is repeated to capture a number of frame pairs. We calculate autocorrelation of all the first frame in each pair and take the average autocorrelation for PR algorithm to derive the fully illuminated object at low resolution (Fig. 4A). Certainly, one can randomly pick a single autocorrelation to derive the fully illuminated object, but the average autocorrelation improves the SNR significantly. For each frame pair, we estimate PSF by deconvolution of the first frame with the reconstructed object (Fig. 4B), then the speckly illuminated object is achieved by deconvolution from the second frame and the estimated PSF. Thanks to deconvolution, we can extract very good information of the speckly illuminated object even with very low correlation of the dynamic scattering media between two shots (Fig. 4C). Our simulation with 20dB noise in image acquisition shows that the main feature of the image is still recovered when the media correlation is only 0.16.

500 speckly illuminated object's images derived from the 500 frame pairs is used to calculate the high order cumulants. Figure 4D presents the fourth order cumulant ($C_4$) image with different media's correlation value. We can reconstruct super-resolution image non-invasively through scattering media with small correlation value between 2 consecutive shots as low as 0.16. The image quality is very much similar for all correlation values higher than 0.16. The artefacts dominate when correlation is 0.03, and high order cumulants cannot suppress it. The normalized intensity profiles across the line marked by dash arrows are plotted in Fig. 4E, illustrating the super resolution imaging through dynamic scattering media of our adaptive NISSFI approach.

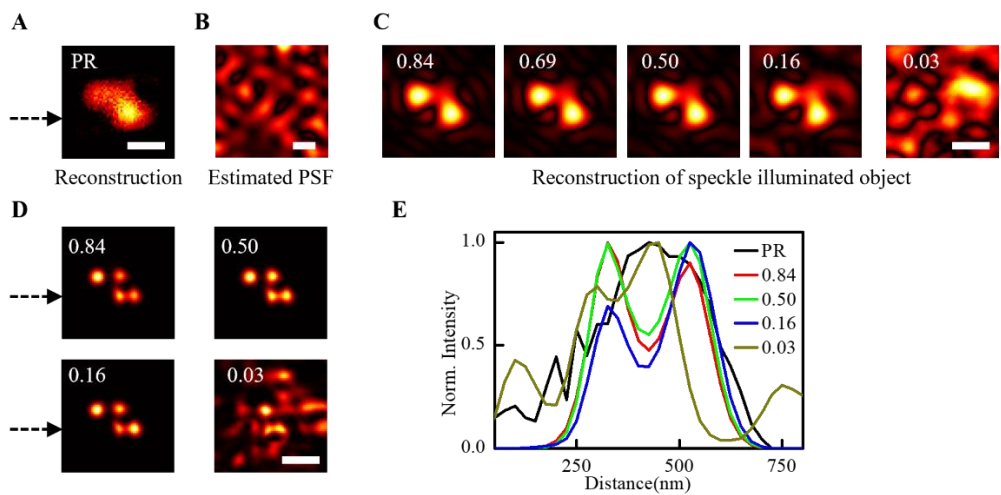

**Fig. 4. Simulation for NISSFI through dynamic scattering media.** (**A**) Reconstruction of full illuminated object. (**B**) Using deconvolution 1 between reconstruction and a full speckle frame to estimate PSF, then using deconvolution 2 between estimated PSF and fluctuating decaying speckle frames can yield a series of fluctuating frames. (**C**) The correlation is decreasing gradually for dynamic scattering media. (**D**) Comparing fluctuating frames between dynamic scattering media (first row) and ground truth(second row). (**E**) The $C_4$ from dynamic scattering media with different correlation, the first row is slow case, the second row is extremely case. (**F**) The intensity profile along the line indicated by dashed arrow in (**A**) and (**E**). (**G**) The same as (**F**) but for extremely case. Scale bar: 500 nm.

### NISSFI with fluctuating emitters

When illuminating a transmission sample with random speckle patterns as presented above, small illuminating speckle sizes are desired so that the different dots on the sample fluctuate independently from each other. If the distance between two adjacent dots is smaller than the illumination speckle size, their fluctuation will have some correlation, and the higher cumulant might not resolve them well. However, the smallest illumination speckle size is also diffraction-limited, as defined by the distance from a sample to the first scattering media and the scattering angle. Therefore, NISSFI with speckle illumination cannot resolve 2 dots with inter-distance less than the diffraction limit of illumination speckle size, though the FWHM of a well-separated point can be scaled down as small as possible by $\sqrt{N}$, where N is the cumulant order.

In fact, the above-mentioned limitation does not exist if the sample comprises of fluctuating emitters (quantum emitters or fluorescent molecules), whose emission intensity fluctuates independently from each other even under the same excitation. We demonstrate this experimentally in a slightly different way with the same sample in Fig. 3. The sample is illuminated uniformly (by enlarging the illumination source as above) with randomly varied intensities. The camera captures 100 speckle frames which are theoretically similar to each other except for different intensities. Then the process is repeated after linearly shifting the sample 500 nm in the vertical direction by a piezo stage to create another set of 100 speckle frames. We randomly pick 2 frames, one from each set, and super-pose them to create a single speckle frame, which would be considered as the speckle frame if the sample was simultaneously presented at two slightly different positions (Fig. 5A). Indeed, the intensities of the vertically adjacent dots are fluctuating independently from each other, while the intensities of two horizontally adjacent dots on the right are fluctuating in absolute correlation.

From 500 speckle frames, we apply our NISSFI method and achieve significant resolution enhancement. With the diffraction limits of 875 nm (NA = 0.37), the PR result cannot resolve adjacent 400-nm dots with center-to-center distance 500 nm, i.e., only 100 nm gap between them (Fig. 5B). Nevertheless, it helps us to estimate the PSF as presented in Fig. 5C. Figure 5D presents our NISSFI results where the higher order cumulants gradually show 2 vertically separated dots. The mean and $C_2$ images cannot resolve the 100nm vertical gap between two dots, but $C_4$ and especially $C_8$ images can clearly show the gap. The $32^{th}$ order cumulant result even shows 2 dots as two separated pixels. The intensity profile across the dots' center confirms the resolution enhancement with a clear dip between two peaks at higher-order cumulants. It is understandable that we cannot resolve the two horizontally adjacent dots by NISSFI in this experiment because they do not independently fluctuate.

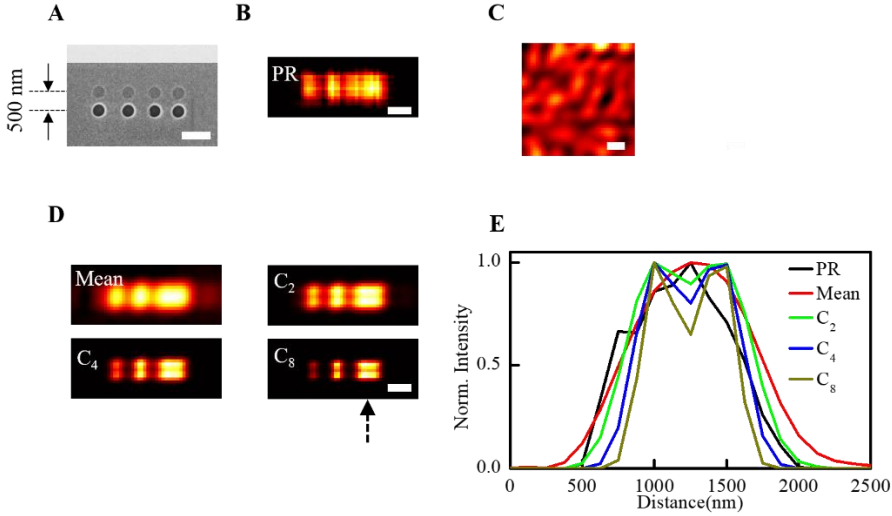

**Fig. 5. NISSFI with artificial fluctuating emitters.** (**A**) The artificial object created by vertically shifting the 4-dot sample by 500 nm. (**B**) PR result from the averaging speckle frame. (**C**) Estimated PSF. (**D**) Reconstructed images by NISSFI at different cumulant orders. (**E**) The intensity profile along the line indicated by dashed arrow in (D). Scale bar: 1.0 μm .

**Discussion**

NISSFI consists of multiple steps, the most critical of which is the reconstruction of low-resolution image of a hidden object by PR. To enhance SNR for the input of PR algorithm, we always utilize averaging strategy, which is the averaging speckle frame for static scattering media and the averaging autocorrelation for dynamic scattering media. In addition, the total number of speckles should be large enough to guarantee the autocorrelation of a finite speckle image is approximately equivalent to object's autocorrelation. Previous studies show that ~1 x $10^6$ speckle grains will have a satisfactory result (*3, 21*). However, in order to break the diffraction limit by high order cumulants, we will need to significantly oversample the speckle size, leading to reduced number of speckles on a finite imaging sensor. Therefore, there is a trade-off between approximate autocorrelation and resolution improvement. The sparsity of the sample also relaxes the requirement of a large speckle number. In the present study, the number of speckles (sensor pixel number /speckle size) is ~ 2.2 x $10^4$ and 8.5 x $10^4$ in simulation and experiment. A camera with more pixels and smaller pixel size will be useful in practical experiments.

Our NISSFI inherits the behaviors of conventional SOFI technique. The higher order cumulants is known for emphasizing the bright pixels while suppressing lower intensity pixels due to the non-linear nature of the technique. On one hand, it helps us to suppress the artifacts, which is obviously lower intensity, but on the other hand, it can distort the object if there is any intensity difference. Our prepared sample with equal intensity (i.e. a binary sample) makes the technique easier to succeed. In addition, we use auto cumulants with zero time lag ($\tau = 0$) to simplify calculation in our current simulation and experiment; other higher order cumulants algorithms like cross-cumulants (*22*), bSOFI(*23*), fSOFI(*24*) could also be suitable for imaging through scattering media.

In summary, we present the non-invasive super-resolution imaging through scattering media based on speckle fluctuation. As long as the intensity of a nanoscale object fluctuates in a random fashion spatially and temporally, we capture multiple the fluctuating speckle frames. Our post-processing algorithm reconstruct the diffraction-limited images of the fluctuating object and super-resolution image is achieved by calculating the higher cumulants. The resolution can be enhanced by factor of $\sqrt{N}$, that can easily break the diffraction-limit by simply calculating higher order cumulant (N). With NA = 0.37 (diffraction limited of 875nm), we can resolve two adjacent 400-nm dots with center-to-center distance 500 nm, i.e., only 100 nm gap between them.

**Materials and Methods**

### Simulation parameters

The simulation frames are generated with 25 nm/pixel. The NA of imaging system is 0.95, making the diffraction-limit (also minimum PSF's speckle size) of 342 nm, corresponding to about 13.7 pixels. With these simulation parameters, we can enhance the resolution from 13.7 pixels to a single pixel for a single dot. There are 1000 fluctuating frames and each frame is 2048 x 2048 pixels.

### Experiment setup

For the illumination part, the pseudo-thermal point source, which is created by focusing 532 nm laser beam on a rotating diffuser (Thorlabs, DG10-1500), is highly scattered by scattering media 1 (5 mm thickness polyurethane foam). The scattering media 1 is almost touching the object to ensure as smaller illumination speckle grains as possible. For the imaging part, the scattering media 2 is a translucent tape (Scotch Magic tape) and ~1 mm away from the object. A microscope objective (Nikon, 40X/0.6NA, WD 2.6-3.2 mm) is used to collect more light into a CCD (Andor iKon-L, 2048 x 2048 pixels). Adjusting the distance between scattering media 2 and object to get as many as speckle numbers in CCD. The CCD is placed at about 30 cm from the microscope objective to not only have adequate magnification to image a nanoscale object but also collect a sufficient number of speckles for the computational algorithm. Though the magnification in our setup corresponds to the sampling of 125 nm/pixel on the object plane, the nature of non-invasive imaging does not provide the magnification because of the unknown distance from the scattering media to the sample. In fact, we can always resolve the object by viewing angles from scattering media's view.

We would like to note here that we use an infinite corrected microscope objective without a tube lens. The light after the objective goes directly to the camera. The objective is simply to collect more light into the camera. When we remove the scattering media and image the sample directly, we adjust the objective to achieve the sharpest (smallest spot) on the camera. Obviously, our imaging setup has NA= 0.37, which is smaller than the objective NA of 0.6 designed with a tube lens in a standard microscope. We verify this point in Supplementary Fig. S3.

### Nanoscale object fabrication

The transmission nanoscale objects are fabricated by depositing a 250-nm-thick gold film on ITO glass, then milling various shapes using focus ion beam (FIB) technique.

### Data processing

The center 1536 x 1536 pixels from the averaged frame are used for PR algorithm (*3*). For simulation, we use the whole 2048 x 2048 pixels. The PR reconstruction image are centered and rotated to the correct direction tentatively. Padding zeros to the reconstructed image to the corresponding speckle frames size before estimating PSF.

Deconvolution method is the standard Wiener-deconvolution as presented in Eq. (1), and the speckle frames utilized are raw data without processing.

$$A = F^{-1}\left[\frac{F(I)F(B)^c}{|F(B)|^2 + \alpha}\right] \quad (1)$$

Where $I$ is the speckle frames, $F$ ($F^{-1}$) is the Fourier transform and its inversion, respectively, $(.)^c$ is the complex conjugate, and $\alpha$ is noise to signal ratio. To estimate PSF, $A$ and $B$ are estimated PSF and PR reconstructed image; to calculate the fluctuating object, $A$ and $B$ are fluctuating object and the estimated PSF.

The higher order cumulant calculation is auto-cumulants as Eq. (2) shows, and time lag τ is set to zero for simplifying computation.

$$C_n = G_n - \sum_{i=1}^{n-1}\binom{n-1}{i}C_{n-i}G_i \quad (2)$$

where $C_n$ and $G_n$ are $N_{th}$ order cumulants and $N_{th}$ order correlations respectively. We suggest referring to previous studies (*17, 25*) for more details.

The FWHM is calculated from Gaussian fitting curves.

**Acknowledgments:**

**General**: The authors especially thank Professor Nikolay I. Zheludev at the University of Southampton and Nanyang Technological University, Singapore for great discussions and providing suggestions for improvement. We would like to thank the Ministry of Education – Singapore (MOE), Nanyang Technological University Singapore (NTU), and IIT Goa's start-up grant for their financial support.

**Funding:** The Ministry of Education – Singapore (MOE): MOE-AcRF Tier-1 (MOE2019-T1-002-087), MOE-AcRF Tier-3 (MOE2016-T3-1-006 (S)), Nanyang Technological University Singapore (NTU), IIT Goa's start-up grant (2019/SG/SKS/014), and Science & Engineering Research Board (SERB) (CRD/2020/000311).

**Author contributions:** C.D. initiated the idea. C.D. and S.K.S. supervised the research. X.Z. did all the simulation and experiments based on the initial Matlab code from S.K.S.. C.D., X.Z. and S.K.S. designed the experiments. G.A. fabricated nanoscale samples. L.Y.M.T and D.H.Z. fabricated initial samples to do proof-of-concept experiments then helped to design the final samples. C.D. and X.Z. wrote the manuscript with S.K.S.'s contributions. All authors discussed, analyzed, took responsibility for the results and revised the manuscript.

**Competing interests:** The authors declare that they have no competing interests.

**Data and materials availability:** The sample data for the findings of this study are available upon request.

## Supplementary Materials

Supplementary Materials is available at www.

# Supplementary Materials for

## Non-invasive super-resolution imaging through scattering media using fluctuating speckles

Xiangwen Zhu,[1] Sujit Kumar Sahoo, [2†] Giorgio Adamo,[3] Landobasa Y.M. Tobing,[1] Dao Hua Zhang[1] and Cuong Dang[1†]

*Corresponding author. Email: sujit@iitgoa.ac.in (S.K.S); HCDang@ntu.edu.sg (C.D.)

**This PDF file includes:**

Figs. S1 to S3

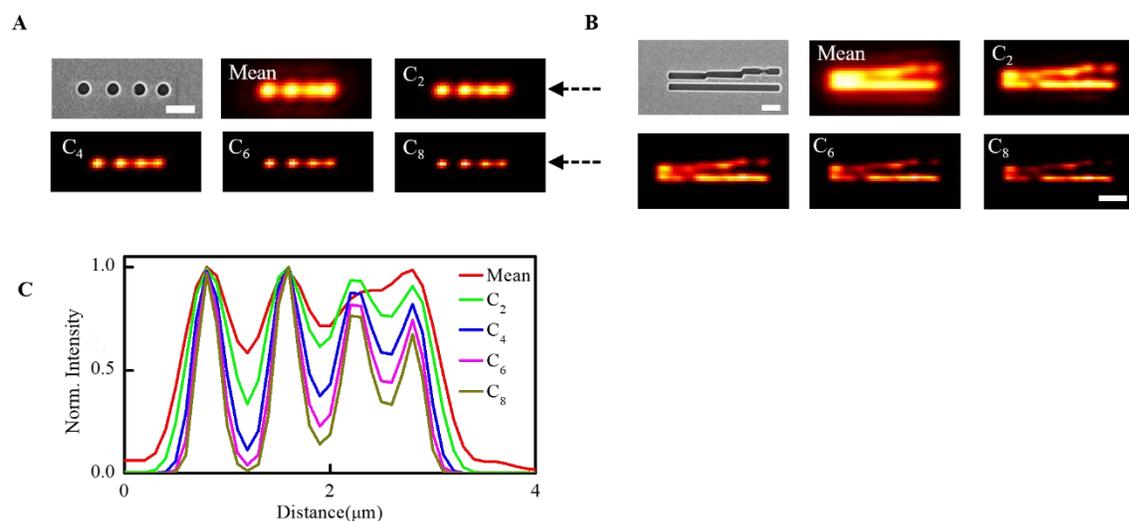

**Fig. S1. Experiment results of conventional SOFI without scattering media.** (**A**) SEM image and reconstruction images of the dot object at different cumulant orders. (**B**) SEM image and reconstruction images of separated lines with different gaps. (**C**) The cross-section indicated by dashed arrow in (**A**). Scale bar: 1.0 μm.

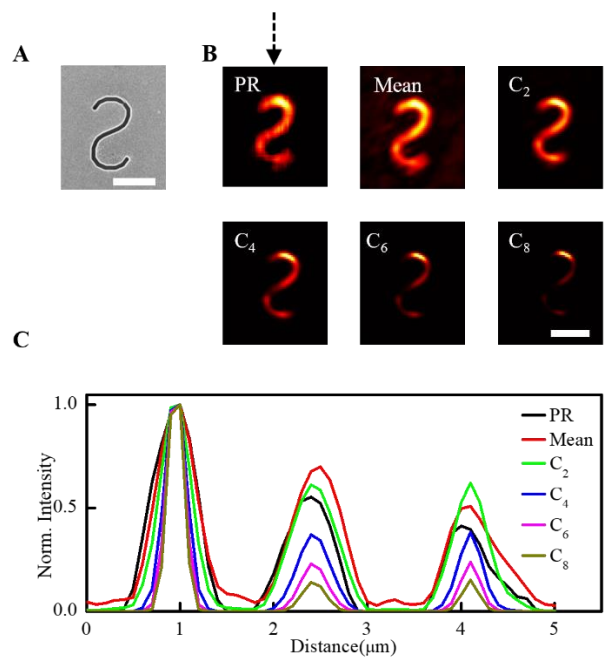

**Fig. S2. Super-resolution imaging for a nanoscale object.** (**A**) The SEM image of nanoscale object (~ 4 μm x 2 μm with line width of 200 nm). (**B**) Reconstruction images. (**C**) The intensity profile along the line indicated by the dashed arrow in (**B**). Scale bar: 2.5 μm.

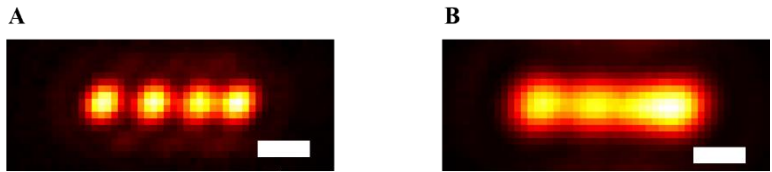

**Fig. S3:** The diffraction limit of the infinite-corrected microscope objective in the setup with a tube lens (**A**) and without a tube lens (**B**).